\documentclass[prb,preprint,tightenlines]{revtex4-1}

\usepackage{amsmath}

\usepackage{graphicx}

\begin{document}

\title{The quantum Gaussian well}
\author{Saikat Nandi}
\email{saikat@tifr.res.in}
\affiliation{Tata Institute of Fundamental Research, Mumbai-400005, India}


\begin{abstract}
Different features of a potential in the form of a Gaussian well have been discussed extensively. Although the details of the calculation are involved, the general approach uses a variational method and WKB approximation, techniques which should be familiar to advanced undergraduates. A numerical solution of the Schr\"{o}dinger equation through diagonalization has been developed in a self-contained way, and physical applications of the potential are mentioned.
\end{abstract}

\maketitle

\section{Introduction}
Potentials such as the infinite square well, the harmonic oscillator, the delta function well, and the finite square well are frequently discussed in textbooks\cite{griffiths,merzbacher} as examples that have bound states. In this paper we consider solutions of the time independent Schr\"{o}dinger equation\cite{griffiths}
\begin{equation}
\label{eq:sch_tind}
-\frac{\hbar^2}{2m}\frac{d^2}{dx^2}\psi(x)+V(x)\psi(x)=E\psi(x),
\end{equation}
for the Gaussian well given by
\begin{equation}
\label{eq:potential}
V(x)=-V_0e^{-\alpha x^2} \qquad (-\infty\leq x\leq +\infty),
\end{equation}
where $V_0>0$ and $\alpha>0$. In Eq.~\eqref{eq:sch_tind} $m$ is the mass of the particle and $\psi(x)$ is the eigenfunction. We will obtain an estimate of the ground state energy from a simple variational method and determine that there are a finite number of bound states using the WKB approximation. We also formulate Eq.~\eqref{eq:sch_tind} as a matrix eigenvalue problem, which can be used for any Hamiltonian. For simplicity, we choose units such that $\hbar=m=1$ in the following.

\section{Bound State Criterion}
Consider the Hamiltonian $H$ for a particle in one dimension subjected to a potential $V(x)$,
\begin{equation}
H=-\frac{1}{2}\frac{d^2}{dx^2}+V(x).
\end{equation}
To demonstrate that $H$ possesses a bound state, it is sufficient to construct a real normalized trial function $\psi(x)$ such that the expectation value of $H$ for this trial function is negative, that is, \begin{equation}
\langle H\rangle\equiv \!\int_{-\infty}^{+\infty}\psi H\psi dx<0.
\end{equation}
This sufficiency condition becomes
\begin{equation}
\label{eq:var}
\langle H\rangle=\int_{-\infty}^{+\infty}\left[\frac{1}{2}\Big(\frac{d\psi}{dx} \Big)^2+V\psi^2\right]dx.
\end{equation}
Defining $L$ to be the width (or, more precisely the length scale) of $\psi(x)$, we can write $\psi$ in the form
\begin{equation}
\psi(x)=\frac{1}{\sqrt{L}}\phi\Big(\frac{x}{L} \Big).
\end{equation}
We substitute this form for $\psi$ in Eq.~(\ref{eq:var}) and obtain
\begin{equation}
\label{eq:condition}
\langle H\rangle=\frac{a_1}{2L^2}+\frac{a_2}{L},
\end{equation}
where $a_1=\!\int_{-\infty}^{+\infty}(d\phi/dx')^2dx'$ (with $x'=x/L$), is a dimensionless positive number and $a_2=\!\int_{-\infty}^{+\infty}V\phi^2dx$. If $L$ is sufficiently large the second term on the right-hand side of Eq.~(\ref{eq:condition}) dominates the first term, and $[\phi(x/L)]^2$ becomes $\sim[\phi(0)]^2$ (a positive constant), so that, it can be taken out of the integral. Hence, the sign of $a_2$ depends only on the sign of $\int_{-\infty}^{+\infty}Vdx$. For both $a_1$ and $a_2$ positive, it is not possible to obtain a positive value of $L$ for which $\langle H\rangle$ is negative. Thus, the condition on $V(x)$ to have at least one bound state is
\begin{equation}
\label{eq:bcondi}
\int_{-\infty}^{+\infty}Vdx<0,
\end{equation}
which depends on the shape of the potential, not on its strength.

Equation~\eqref{eq:bcondi} is a sufficient condition for a potential to have a bound state. It is not a necessary condition as the example of the harmonic oscillator potential shows. Note that this general criterion for the existence of the bound state follows from a simple argument involving dimensional analysis and the variational principle.

\section{Ground State Energy from the Variational Method}
In practice, we can obtain various upper bounds of the ground state energy by calculating $\langle H\rangle$ for suitably chosen trial functions. There is a tradeoff between improving the approximate ground state energy using a complex trial wave function and the ease of calculation. Here we choose a normalized Gaussian trial wave function $\psi(x)$ with adjustable width,
\begin{equation}
\psi(x)=\Big(\frac{2b}{\pi} \Big)^{1/4}e^{-bx^2},
\end{equation}
where $b$ is related to the width $L$. The expectation value of the Hamiltonian $H$ becomes,
\begin{equation}
\label{eq:expect}
\langle H\rangle=\frac{b}{2}-V_0\sqrt{\frac{2b}{2b+\alpha}}.
\end{equation}
From Eq.~\eqref{eq:expect} the condition for a minimum is $d \langle H\rangle/db=0$,
or
\begin{equation}
\label{eq:energy}
b(2b+\alpha)^3 =2V_0^2\alpha^2.
\end{equation}
\begin{table}[ht]
\centering
\begin{tabular}{|c|c|c|c|c|}
\hline
$V_0$ & $\alpha$ & $b$ & $\langle H\rangle$ & $E_{\rm num}$ \\
\hline
1.0 & 1.0 & 0.3742 & -0.4671 & $-0.4774$ \\
\hline
2.5 & 0.5 & 0.6113 & -1.8005 & $-1.8038$ \\
\hline
3.0 & 1.0 & 0.8717 & -1.9557 & $-1.9637$ \\
\hline
3.0 & 0.1 & 0.3504 & -2.6312 & $-2.6316$ \\
\hline
\end{tabular}
\caption{\label{tab:ground_state}Comparison of the ground state energy $\langle H\rangle$ calculated from
the variational principle and the numerical solution ($E_{\rm num})$. $\langle H\rangle$ is within 2\% of the corresponding numerically evaluated value.}
\end{table}

Equation~\eqref{eq:energy} determines the variational parameter $b$ for given values of $V_0$ and $\alpha$. The variational method ground state energies for different values of $V_0$ and $\alpha$ are given in Table~\ref{tab:ground_state}. As expected, these estimates are greater than the exact (numerically obtained) ground state energies, but are remarkably close.

\section{Finite Number of Bound states}
The wave function $\psi_n(x)$ corresponding to the eigenvalue $E_n$ of a discrete spectrum has $n$ nodes. The nature of the number of bound states is best demonstrated by the WKB approximation. States belonging to the discrete energy spectrum are semi-classical only for large values of $n$. In one dimension the WKB integral for the energy $E$ is given by\cite{merzbacher}
\begin{equation}
\label{eq:wkb}
\int_{x_1}^{x_2}\!\sqrt{2[E-V(x)]}\,dx=\Big(n-\frac{1}{2} \Big)\pi, \quad \text{for $n=1,2,3 \ldots$}
\end{equation}
where $x_1$ and $x_2$ are the turning points of the classical motion. Because the discrete spectrum lies in the range of energy values for which the particle cannot move to infinity, the energy must be less than the limiting values $V(x=\pm\infty)$. For the Gaussian potential this condition implies that $E<0$. Thus, the number of discrete levels $N$ is obtained by setting $E=0$ in Eq.~(\ref{eq:wkb}), so that $n=N$ is close to the quantum number for the last bound state. For the classical turning points $x=\pm\infty$, Eq.~(\ref{eq:wkb}) becomes
\begin{equation}
\label{eq:bound}
\sqrt{2V_0}\!\int_{-\infty}^{+\infty}e^{-\alpha x^2/2}\,dx=\left(N-\frac{1}{2}\right)\pi,
\end{equation}
which gives
\begin{equation}
\label{this formula}
N=\frac{2}{\sqrt{\pi}}\sqrt{\frac{V_0}{\alpha}}+\frac{1}{2}.
\end{equation}
\begin{table}[ht]
\centering
\begin{tabular}{|c|c|c|c|c|}
\hline
$\frac{V_0}{\alpha}$ & $N$ & $N_{\rm num}$ \\
\hline
0.5 & 1.3 & 1 \\
\hline
1.0 & 1.6 & 1 \\
\hline
10.0 & 4.1 & 4 \\
\hline
100.0 & 11.8 & 11 \\
\hline
\end{tabular}
\caption{\label{tab:number_state}Number of bound state given by the WKB approximation ($N$) and the numerical method ($N_{\rm num}$).}
\end{table}
For $V_0$ and $\alpha$ finite, $N$ is also finite. Therefore, the number of bound states is finite for the Gaussian well. The value of $N$ calculated from Eq.~\eqref{this formula} compared to the numerical solution is in Table~\ref{tab:number_state}.
Equation~\eqref{this formula} implies that the number of bound states is a function of the ratio $V_0/\alpha$, which is a measure of the length scale $L$ of the wavefunctions.\cite{note1} Because $V(x)\propto(\sqrt{\alpha/\pi})e^{-\alpha x^2}$, $V(x)$ becomes a delta function in the limit $\alpha\rightarrow\infty$, for which Eq.~(\ref{eq:bound}) gives $N=1/2$, implying that there exists exactly one bound state for the delta function well.\cite{griffiths} The reason for the half-quantum number is that the ground state has no nodes. If $V_0\rightarrow\infty$, then $N\rightarrow\infty$, which is the case for an infinite square well.\cite{griffiths}

\section{Numerical Solution of the Schr\"{o}dinger Equation}
The one-dimensional Schr\"{o}dinger equation has few analytic solutions, and most problems must be solved numerically. We first use a Taylor's series to obtain a discretization of derivatives for a function $f(x)$ and write the second derivative of the function $f(x)$ in the three point central difference form
\begin{equation}
\label{eq:discrete}
f''(x)=\frac{f(x+\delta)-2f(x)+f(x-\delta)}{\delta^2}+O(\delta^2),
\end{equation}
where, $\delta$ is the step size. If $r$ is the number of mesh points and $x_{\max}$, $x_{\min}$ are the maximum and minimum value of the variable $x$, the step size $\delta$ is
\begin{equation}
\delta=\frac{x_{\max}-x_{\min}}{r}.
\end{equation}

The solution of Eq.~\eqref{eq:sch_tind} is accomplished by discretizing it using Eq.~(\ref{eq:discrete}) and evaluating the functions and derivatives at $x_k=x_{\min}+k\delta$ (for $k=1,2, \ldots,r-1$),
\begin{equation}
\label{eq:sch_dis}
-\frac{\psi(x_k+\delta)-2\psi(x_k)+\psi(x_k-\delta)}{2\delta^2}+V(x_k)\psi(x_k)=E\psi(x_k),
\end{equation}
Because the solution is expected to decay exponentially outside $x_{\max}$ and $x_{\min}$, we solve Eq.~\eqref{eq:sch_dis} in the interval $[x_{\max},x_{\min}]$. These cut-offs and the step size need to be adjusted to obtain the desired accuracy.

Equation~(\ref{eq:sch_dis}) is equivalent to the tight-binding approximation applied to a chain of atoms with spacing $\delta$ and one orbital per atom.\cite{ajp} We can develop a matrix representation of the Schr\"{o}dinger equation, with $-\delta^{-2}/2$ in the sub-diagonal and super-diagonal matrix elements and the diagonal elements are $\delta^{-2}+V(x_k)$. Equation~(\ref{eq:sch_dis}) can then be written as a matrix equation as,
\begin{eqnarray}
\label{matrix}
&&\begin{pmatrix}
\delta^{-2}+V(x_1) & -\delta^{-2}/2 & 0 & 0 & \ldots & 0 & 0\\
-\delta^{-2}/2 & \delta^{-2}+V(x_2) & -\delta^{-2}/2 & 0 & \ldots & 0 & 0\\
\ldots & \ldots & \ldots & \ldots & \ldots & \ldots & \ldots\\
0 & \ldots & \ldots & \ldots & \ldots & \delta^{-2}+V(x_{r-2}) & -\delta^{-2}/2\\
0 & \ldots & \ldots & \ldots & \ldots & -\delta^{-2}/2 & \delta^{-2}+V(x_{r-1})
\end{pmatrix}
\begin{pmatrix}
\psi(x_1)\\
\psi(x_2)\\
\vdots\\
\psi(x_{r-2})\\
\psi(x_{r-1})\\
\end{pmatrix} \nonumber \\
&& \quad=E
\begin{pmatrix}
\psi(x_1)\\
\psi(x_2)\\
\vdots\\
\psi(x_{r-2})\\
\psi(x_{r-1})\\
\end{pmatrix}
\end{eqnarray}

Equation~\eqref{matrix} is a matrix eigenvalue problem with a tridiagonal matrix of dimension $(r-1)\times(r-1)$; thus there are $(r-1)$ eigenvalues. Because all the matrix elements are real and the transpose of the matrix is equal to the matrix itself, it is a Hermitian matrix, and hence all the eigenvalues are real. An efficient way of diagonalizing a tridiagonal matrix is to use the standard LAPACK routine DSTEVX,\cite{lapack} which stores the symmetric tridiagonal matrix in two one-dimensional arrays, one of length $(r-1)$ containing the diagonal elements, and one of length $(r-2)$ containing the off-diagonal elements, and returns the eigenvalues along with the eigenfunctions of the matrix.
\begin{figure}[h]
\includegraphics{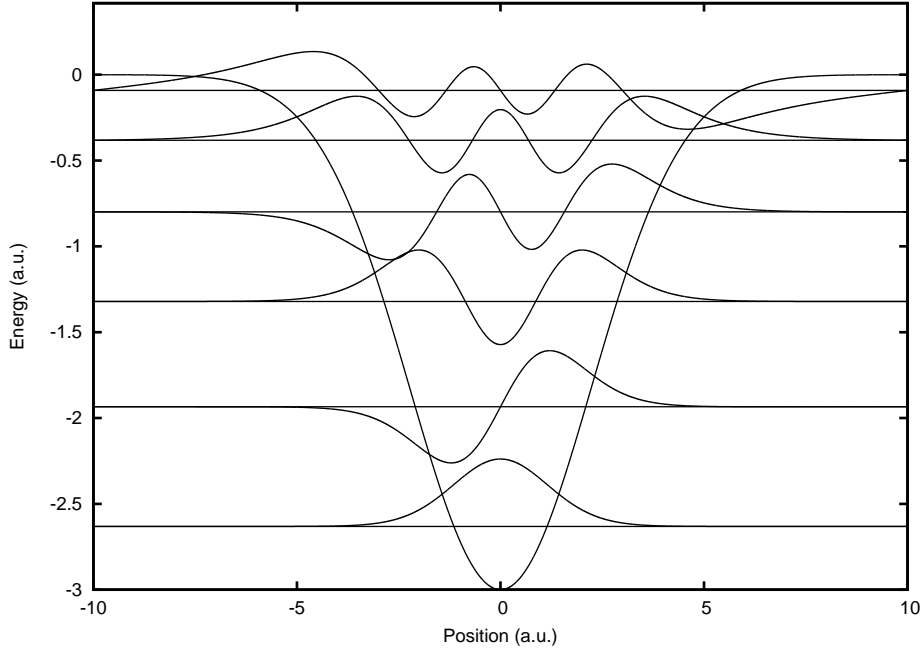}
\caption{\label{fig:wavefunction}The energy eigenfunctions and the corresponding eigenvalues for $V_0=3.0$
and $\alpha=0.1$. Note the unequal spacing between different levels.}
\end{figure}

The eigenvalues and corresponding eigenfunctions for the Gaussian well with $V_0=3.0$ and $\alpha=0.1$ are shown in Fig.~\ref{fig:wavefunction}. The ground state is symmetric with respect to the center of the potential (even parity), and the first excited state is antisymmetric (odd parity). The wave functions resemble those of the harmonic oscillator,\cite{griffiths} because for $x<1$, the dominant term in the expansion of $e^{-\alpha x^2}$ is proportional to $x^2$.

The virtue of the matrix method is that it can be applied to any potential for which the Hamiltonian can be brought into a symmetric tridiagonal or bidiagonal matrix. However, for systems with periodic boundary conditions, the Hamiltonian is no longer tridiagonal and we cannot use the simple matrix method.

\section{Tunneling In a Gaussian Barrier}
A Gaussian barrier can be constructed by changing the sign of $V_0$ in Eq.~(\ref{eq:potential}). Because the condition in Eq.~(\ref{eq:bcondi}) is not satisfied, there is no bound state. The barrier formed in this way is an interesting example of tunneling. We start with the time energy uncertainty principle
\begin{equation}
\label{prev1}
\Delta E\Delta t\simeq\frac{1}{2}.
\end{equation}
Denote the energy of the incident particle by $E$. The uncertainty in the energy is $\Delta E$ and for sufficiently small $\Delta t$ the energy of the particle $E+\Delta E$ is greater than the height of the barrier $V_0$. Tunneling will take place if in the time $\Delta t$ the particle can traverse the barrier. We take $\alpha^{-1/2}$ as the width of the barrier and write,
\begin{equation}
\label{prev2}
\Delta t\simeq\frac{\alpha^{-1/2}}{\sqrt{2(E+\Delta E-V_0)}}.
\end{equation}
From Eqs.~\eqref{prev1} and \eqref{prev2} we find $\Delta E$ to satisfy the equation,
\begin{equation}
(\Delta E)^2-\frac{\alpha}{2}\Delta E+\frac{\alpha}{2}(V_0-E)=0.
\end{equation}
The condition for $\Delta E$ to be real implies that
\begin{equation}
\frac{\alpha}{8}>(V_0-E), \label{left}
\end{equation}
which is the condition for tunneling. The left-hand side of Eq.~\eqref{left} is the kinetic energy of the particle obtained from the `position-momentum uncertainty relation, with an uncertainty of $\Delta x\sim \alpha^{-1/2}$ in the particle's position. We see that for tunneling to occur, the kinetic energy of the particle must be greater than the difference between the height of the barrier $V_0$ and the total energy $E$.

We now investigate tunneling using the WKB approximation. Consider the classically inaccessible region, $E<V(x)$, as $a\leq x\leq b$, so that we can write the transmission coefficient as\cite{merzbacher}
\begin{equation}
\label{eq:trans}
T\approx \exp\left[-2\!\int_a^b \kappa(x)dx\right],
\end{equation}
where $\kappa(x)=\sqrt{2(V(x)-E)}$. We define the opacity of the barrier by
\begin{equation}
\label{eq:coeff}
\theta=\exp\left[\int_a^b \kappa(x)dx\right],
\end{equation}
so that Eq.~(\ref{eq:trans}) becomes $T\approx \theta^{-2}$. No attempt been made to obtain an analytical solution to the integral in Eq.~(\ref{eq:trans}). But we can find an approximate solution of the integral that correctly predicts the transmission coefficient $T$.

We introduce the dimensionless quantity $\beta= V_0/E>1$ and write the classical turning points as $a=-\sqrt{\ln \beta/\alpha}$ and $b=\sqrt{ \ln \beta/\alpha}$. The change in variables, $x=y\sqrt{\ln\beta/\alpha}$, changes Eq.~(\ref{eq:coeff}) to
\begin{equation}
\theta=\exp\left(\sqrt{\frac{2V_0 \ln\beta}{\alpha}}\!\int_{-1}^{+1}\sqrt{\left[1-\left(1-\beta^{-y^2}+\beta^{-1}\right)\right]}dy\right).
\end{equation}
\begin{figure}[h]
\includegraphics{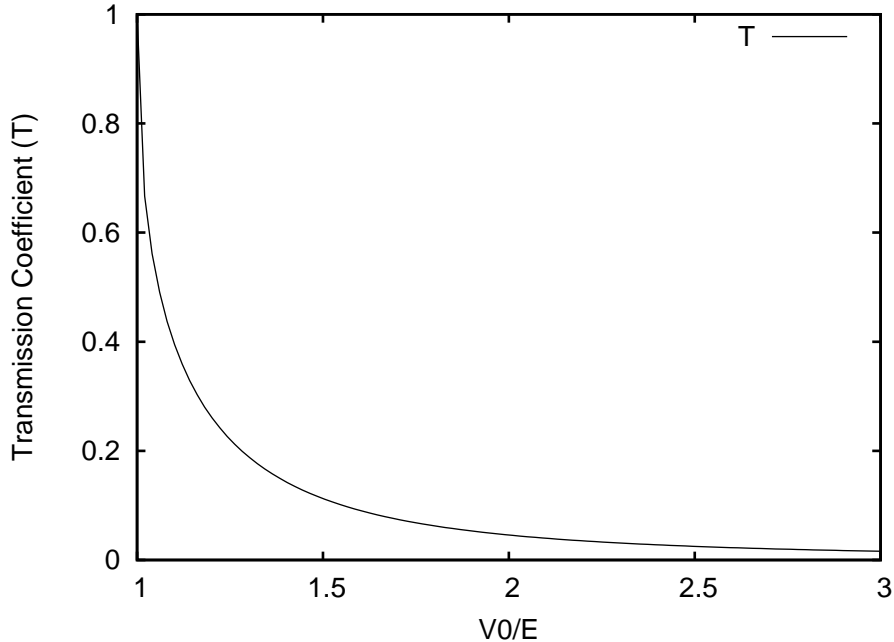}
\caption{\label{fig:transmission}Dependence of the tansmission coefficient $T$ on $\beta=V_0/E$. As expected, $T$ decreases as $\beta$ increases.}
\end{figure}

The term $(1-\beta^{-y^2}+ \beta^{-1})$ is always less than unity. If we assume that this term is approximately equal to unity,\cite{note2} then the binomial expansion of $\left[1-\left(1-\beta^{-y^2}+\beta^{-1}\right)\right]^{\frac{1}{2}}$ to first order we have
\begin{equation}
\theta\approx \exp\left(\sqrt{\frac{2V_0 \ln \beta}{\alpha}}\int_{-1}^{+1}\left[\frac{1}{2}+\frac{1}{2}\beta^{-y^2}-\frac{1}{2\beta}\right]dy\right),
\end{equation}
which can be further expressed as
\begin{equation}
\label{eq:theta}
\theta\approx \exp\left(\sqrt{\frac{2V_0}{\alpha}}\left[\sqrt{\ln \beta}+\frac{\sqrt{\pi}}{2}\mbox{erf}\left(\sqrt{\ln\beta}\right)-\frac{\sqrt{\ln \beta}}{\beta}\right]\right),
\end{equation}
where erf(x) is the error function of $x$. Equation~(\ref{eq:theta}) implies that $T$ depends only on the ratio $\beta=V_0/E$, a common feature of barrier tunneling. The dependence of $T$ on $\beta$ can be seen from Fig.~\ref{fig:transmission}. It is evident that the transmission coefficient decreases as $\beta$ increases. As expected for $\beta=1$, that is, $E= V(x=0)$, $T=1$, implying no reflection from the barrier.

\section{Two Simple Application}
Thus far we have discussed different aspects of the potential from a quantum theoretic approach. The Gauusian well (or, barrier) is not a long range potential, because it falls off faster than $\frac{1}{x^2}$. However, it has a crucial advantage over  the widely used `finite square well' potential in that  it is continuous throughout the entire range $[-\infty<x<+\infty]$, whereas the finite square well goes to zero discontinuously. In the following we describe two applications of this potential.

\subsection{Single Particle Motion in Atomic Nuclei}
The `mean-field' dynamics of a single nucleon in the field of all other nucleons is the starting point of nuclear many-body theory. The Gaussian well potential is  well suited for describing the interaction of a nucleon (especially, a neutron) with the heavy nucleus.\cite{flugge} The force between them is appreciable only over a very short distance, of the order of $10^{-15}$m, and in this range the forces are very large compared to forces holding atoms together. Keeping in mind the three dimensionality of a nucleus, we can change our variable from $x$ to the radius vector $\textbf{r}$ in spherical polar coordinates, so that the potential can be represented by a `half' Gaussian well,
\begin{equation}
\label{eq:gauss_half}
V(\textbf{r}) =
\begin{cases}
-V_0e^{-\alpha r^2} & \text{$r\geq 0$} \\
~ 0 & \text{$r < 0$}.
\end{cases}
\end{equation}
It is monotonically increasing with distance (i.e. attracting) and approaches zero very quickly as $r$ goes to infinity, reflecting the short-range nature of the nuclear force. The parameter $\alpha$ can be adjusted to the experimental value of the nuclear interaction barrier. From the Schr\"{o}dinger equation in radial form, one has
\begin{equation}
-\frac{\hbar^2}{2m}\frac{d^2u}{dr^2}+\left[-V_0e^{-\alpha r^2}+\frac{\hbar^2}{2m}\frac{l(l+1)}{r^2}\right]u=Eu,
\end{equation}       
where $l$ is the azimuthal quantum number. It follows immediately that this equation is identical to Eq.~(\ref{eq:sch_tind}) for $l=0$. It is left as an exercise to the reader to solve this equation using the numerical method discussed already and  to deduce the single particle nuclear energy levels for $l=0$ by taking appropriate values of $V_0$ and $\alpha$.\cite{note3} For a potential such as the one in  Eq.~(\ref{eq:gauss_half}) there will be no even-parity states (because at $r=0$ there must be an   antinode.).

\subsection{Scanning Tunneling Microscope}
The Gaussian barrier can also be very useful in analyzing a scanning tunneling microscope (STM). The basic idea behind a STM is quite simple. An atomically sharp metal tip is brought very close ($<5$ \AA) but without physical contact  to a sample surface, and a small bias voltage (3-6 volts) is applied between them. If the distance is small enough, an electron (with an energy 3-6 eV) can tunnel quantum mechanically through the potential barrier developed in between the tip and sample. This gives rise to the tunneling current that is the result of the overlap between wave functions of the tip-atom and the surface-atom. Assume the metal to tip gap is a Gaussian barrier with a height of $V_0=10$~eV $\simeq 0.36$ in atomic units. An electron with energy of $E=5$~eV $\simeq0.18$ in atomic units approaches the surface. Now from Eq.~(\ref{eq:trans}), the transmission coefficient is
\begin{equation}
T\simeq e^{-2.2\sqrt{\frac{V_0}{\alpha}}}
\end{equation}
If the tip is $\frac{1}{\sqrt{\alpha}}=0.15$~nm $\simeq2.83$ in atomic units from the surface, we have $T\simeq0.024$. However, it should be noted that for a large distance between the tip and surface, e.g., if $\frac{1}{\sqrt{\alpha}}=1$~nm $\simeq18.9$ in atomic units,  $T$ will become $\sim10^{-11}$ and practically, there will be no current
 to measure~! Thus the magnitude of the tunneling current is extremely sensitive to the gap between the tip and sample. The position of the tip in three dimensions is accurately controlled by piezoelectric drivers. The tip is scanned in two lateral dimensions, while a feedback circuit constantly adjusts the tip height to keep the current constant. As we measure the current with the tip moving across the surface, information about the atomic nature  of the surface can be determined by tracing the path of the tip.

As a final example there is the  alpha particle tunneling problem which also can be modeled with the Gaussian barrier.

\section{A Double Gaussian Well}
\begin{figure}[h]
\includegraphics{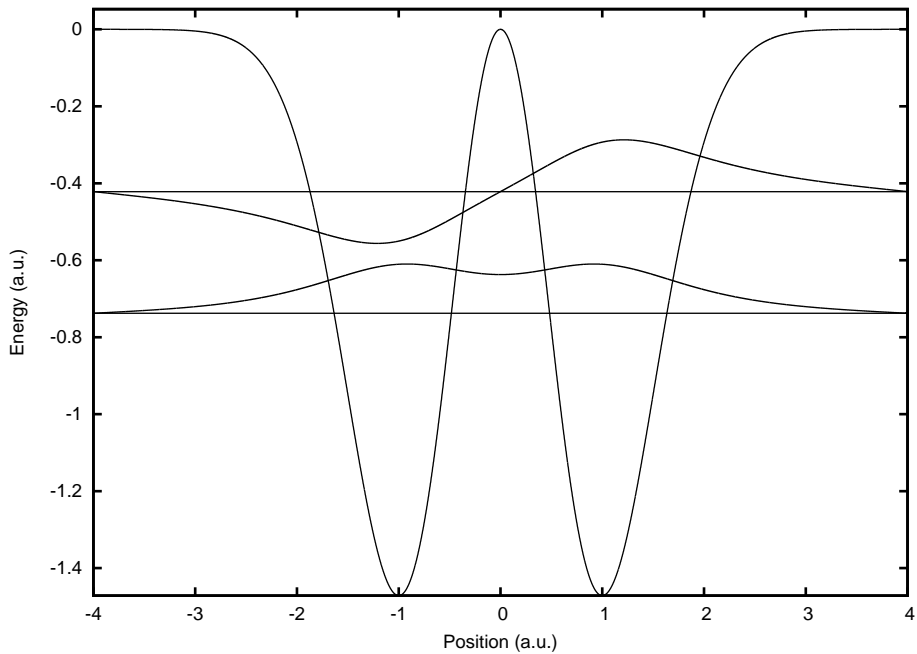}
\caption{\label{fig:double_well1}The energy levels (ground state \& first excited state) along with the wavefunctions for $U(x)$ with $V_0=3.0$ and $\alpha=1.0$. We can see the tunneling of the wavefunction through the barrier.}
\end{figure}

An interesting double well potential may be formed by multiplying $V(x)$ in Eq.~\eqref{eq:potential} by a factor of $x^2$. Because the potential $U(x)=-V_0 x^2 e^{-\alpha x^2}$ is symmetric, the Hamiltonian can be brought into a symmetric tridiagonal form. The numerical solution for this potential is given in Fig.~\ref{fig:double_well1}. Because the barrier is finite the wave function extends into both wells. The ground state wave function is small but nonzero inside the barrier. Assume the particle starts in the right well. From Fig.~\ref{fig:double_well1} we see that initially $\psi=\psi_1+\psi_2$ (where, $\psi_1$ \& $\psi_2$ are the ground state and first excited state, respectively) is canceled on the left. As a consequence the particle will oscillate between the wells with the period, $\tau=2\pi\hbar/(E_2-E_1)$. Thus the tunneling rate depends on the energy difference, $\Delta E=E_2-E_1$, and a double well with a high or wide barrier will have a smaller $\Delta E$ than one with a low or narrow barrier. Also, $\Delta E$ will become larger as the energy increases (that is, as $(V_0-E)$ decreases).

The double Gaussian well is a good model for a two level system. 
\begin{figure}[h]
\includegraphics{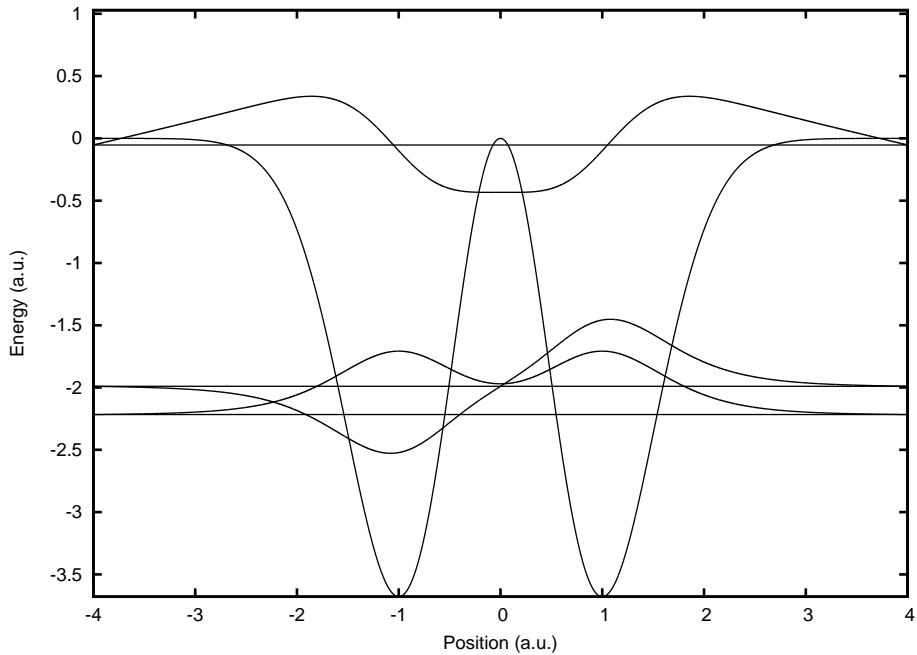}
\caption{\label{fig:double_well2}The wavefunctions and the corresponding energy eigenvalues of $U(x)$ for $V_0=10.0$ and $\alpha=1.0$. The proximity between the lowest two states is evident.}
\end{figure}
An example is a quantum well laser, based on the transition that an electron makes between the ground and first excited state of a double Gaussian well. By choosing $\alpha$ and $V_0$, we can tune the wavelength of the light emitted. 
Another not so obvious case could be a Qubit. The quantum bit, or qubit, is the simplest unit of quantum information. Measurements give only two  values: `zero' or `one'. It is described conveniently by a state in a two level quantum-mechanical system. A pure qubit state would then be a linear superposition of those two states. Though the double Gaussian well has multiple energy levels, it can be seen from Fig.~(\ref{fig:double_well2}) that the relative spacing between the ground and first excited state is very small compared to that between the first and second excited states. Therefore, the lowest two bound states can be effectively decoupled from the other states by choosing appropriate values of $V_0$ \& $\alpha$. An electron making transitions between these two states can easily be  considered as a `charge qubit', a much discussed way for consistent quantum data storage.

\begin{acknowledgments}
The author would like to thank Sayan Chakraborti, Swastik Bhattacharya, and Shamashis Sengupta for many helpful suggestions. He is grateful to Professor Deepak Dhar for help in understanding different aspects of quantum mechanics.
\end{acknowledgments}


\begin{thebibliography}{10}
\bibitem{note1}Here $\sqrt{1/\alpha}$ is the length scale for the potential. Now keeping the ratio $(V_0/\alpha)$ fixed, the length scale of the wavefunctions can be identified as $L \sim \frac{1/\sqrt{\alpha}}{V_0/\alpha}=\sqrt{\alpha}/V_0$.
\bibitem{note2}We can add more terms in the binomial expansion to obtain more accurate values of $T$. However, the essential physical behavior of $T$ will remain the same.
\bibitem{note3}In principle, one can also solve for $l\neq 0$ by this numerical method. The diagonal terms in the tridiagonal matrix will then be modified accordingly.
\end{thebibliography}
\end{document}